# Spatially anisotropic Kondo resonance coupled with the superconducting gap in a kagome metal

Zichen Huang[1,2#], Hui Chen[1,2,3#✉], Zhongqin Zhang[4#], Hao Zhang[1,2#], Zhen Zhao[1,2], Ruwen Wang[1,2], Haitao Yang[1,2,3], Wei Ji[4✉], Ziqiang Wang[5], and Hong-Jun Gao[1,2,3✉]

[1] *Beijing National Center for Condensed Matter Physics and Institute of Physics, Chinese Academy of Sciences, Beijing 100190, PR China*

[2] *School of Physical Sciences, University of Chinese Academy of Sciences, Beijing 100190, PR China*

[3] *Hefei National Laboratory, 230088 Hefei, Anhui, PR China*

[4] *Beijing Key Laboratory of Optoelectronic Functional Materials & Micro-Nano Devices, School of Physics, Renmin University of China, 100872, Beijing, China*

[5] *Department of Physics, Boston College, Chestnut Hill, MA 02467, USA*

[#] These authors contributed equally to this work

✉ Correspondence to: hjgao@iphy.ac.cn; wji@ruc.edu.cn; hchenn04@iphy.ac.cn

**The chromium-based kagome metal $CsCr_3Sb_5$ has garnered significant interest due to its strong electron correlations, intertwined orders and potential for unconventional superconductivity under high pressure. The evolution of magnetic and superconducting interactions as the more frequently studied $CsV_3Sb_5$ is doped to $CsCr_3Sb_5$ remains poorly understood. Here, we demonstrate the emergence of a spatially anisotropic Kondo resonance intertwined with the superconducting gap, enabled by introducing magnetic Cr impurities into the kagome superconductor $CsV_3Sb_5$. The addition of dilute Cr impurities not only weakens long range charge density wave order but also produces local magnetic moments that leads to Kondo resonances. We show that the Kondo resonance forms anisotropic, ripple like spatial patterns around individual Cr atoms, breaking all local mirror symmetries. We further reveal that with the emergence of Kondo screening, the coherence peak and depth of superconducting gap with finite zero-energy conductance are enhanced. This suggests that non superconducting carriers at the Fermi surface in the parent compound participate in the Kondo effect, simultaneously screening Cr magnetic moments and increasing the superfluid density. Our findings offer an opportunity to study the interplay between superconductivity and local magnetism in kagome materials.**

The interplay between magnetism and superconductivity has been a central topic of condensed matter physics. While magnetism generally suppresses phonon mediated conventional superconductors, magnetic fluctuations are widely considered a key pairing mechanism in unconventional superconductors. Antiferromagnetic (AFM) fluctuations are the most common driving force of superconductivity in cuprates, iron-based superconductors and heavy-fermion superconductors[1,2], while ferromagnetic fluctuations has been associated with spin-triplet superconductivity, such as $UTe_2$[3]. Recently, the vanadium-based kagome metal $AV_3Sb_5$ ($A$ = K, Rb, Cs) have drawn sustained interest as kagome systems exhibiting superconductivity[4,5]. $AV_3Sb_5$ possess cascade of exotic phenomena such as multiple Van Hove singularity near the Fermi level[6,7], $Z_2$ nontrivial topology[5], anomalous Hall effect[8], pair density wave[9–11] and electronic nematicity[12–14]. $AV_3Sb_5$ has V-shaped superconductive gap with sign-preserving multi-band superconductivity[15,16], and the superconductivity in this class of material has been proved to be sensitive to external perturbation. Double-dome superconductivity and unconventional competing with CDW are observed upon pressure[17,18]. In addition, Chemical substitution provides an additional tuning parameter, yielding complex phase diagrams with intertwined instabilities, such as hole doping via Ti substitution[19,20] and isovalent substitution with Ta or Nb[21,22]. However, the doping-induced magnetism in kagome superconductors has been rarely explored.

Substituting V with magnetic Cr is expected to introduce magnetism into $CsV_3Sb_5$[23–25]. In addition, the related chromium-based material $CsCr_3Sb_5$ shares the same crystal structure and exhibits an AFM transition under 55 K at ambient pressure. With increasing pressure, AFM order is suppressed and superconductivity emerge at 4.2 GPa near the quantum criticality point[26,27]. The multi-phase diagram of $CsCr_3Sb_5$ resembles that of other USCs, and AFM fluctuations is proposed as pairing mechanism, positioning $CsCr_3Sb_5$ as a potential unconventional superconductor candidate within kagome systems. Clarifying how Cr doping modifies magnetism and superconductivity in $CsV_3Sb_5$ is therefore essential for understanding both AFM order and pressure-induced superconductivity in kagome systems.

Here, we investigate the interaction between local magnetic moments and superconductivity in magnetic Cr-doped kagome superconductor $CsV_{3-x}Cr_xSb_5$ by utilizing ultra-low temperature scanning tunneling microscope/spectroscopy (STM/STS). In addition to the gradual suppression of long-range CDW order, dilute Cr doping induces Kondo resonances near the Fermi level, arising from the screening of local magnetic moments. The resonances propagate preferentially along one of the four equivalent directions, thereby breaking all local mirror symmetries. Density functional theory (DFT) calculations reveal that the V-Cr antiferromagnetic coupling induces magnetic frustration among V sites in the V-Sb kagome layer near Cr dopants, leading to ripple-like propagation of spin density, which contributes to the observed anisotropic Kondo resonance. Simultaneously, superconductivity evolves strikingly with Cr doping. In Kondo resonance phase, the coherence peak height and gap depth of superconducting gap significantly enhances, accompanied by unconventional spatial evolution of vortex bound states. Once the Kondo resonance disappears, the superconductivity is rapidly suppressed by magnetic ordering. These findings reveal a distinct superconducting phase in $CsV_{3-x}Cr_xSb_5$, providing insight into the coupling between superconductivity and magnetism in kagome metals.

We first investigate the topographic characteristics of Cr dopants in $CsV_3Sb_5$, achieved by substituting V atomic with Cr in the V-Sb kagome plane. The concentration of Cr dopants is described by $x$ in the $CsV_{3-x}Cr_xSb_5$ (Fig. 1**a**). We have synthesized series of $CsV_{3-x}Cr_xSb_5$ with $x$ ranging from 0 to 1.2 (Fig. 1**b-g** and Fig. S1). Compared with pristine sample, there are additional dumbbell-like topographic protrusions in the STM images of Sb terminated surfaces (Fig. 1**h**). The density of dumbbell-like protrusions increases simultaneously with the Cr concentrations (Fig. 1**b-g**). We thus attribute the dumbbell-like protrusions to the Cr dopants. The Cr dopants are embedded in the middle of two adjacent topmost Sb atoms, which results in the two-fold dumbbell-like protrusions located halfway between pairs of topmost Sb atoms (Fig. 1**h**). Due to the global symmetry of hexagonal lattice, there are three equivalent two-fold axes of dumbbell-like protrusions, which correlates to three lattice directions (green shades in Fig. 1**a**).

We then study the evolution of CDW orders with Cr concentration by performing Fourier transforms (FT) of topographic images. In pristine $CsV_3Sb_5$, the long-ranged 2×2 CDW ($\boldsymbol{Q}_{3q\text{-}2a}$) and $4a_0$ unidirectional charge order ($\boldsymbol{Q}_{1q\text{-}4a}$) have been observed at the Sb surfaces[9,28]. For diluted doped sample ($x$ = 0.012), the Bragg peaks together with clear signature of $\boldsymbol{Q}_{3q\text{-}2a}$ and $\boldsymbol{Q}_{1q\text{-}4a}$ are observed (Fig. 1**i**). For the $x$=0.3 sample, the intensity of CDW peaks become much weaker (Supplementary Fig. 1**a-f**). Thus, we plot the FFT profile along high symmetry direction of samples, which shows that the peak intensity of both $\boldsymbol{Q}_{3q\text{-}2a}$ and $\boldsymbol{Q}_{1q\text{-}4a}$ gradually reduce with increasing Cr concentrations (Fig. 1**j**). The CDWs are nearly undetectable at a large concentration of $x$=1.2, consistent with transport results showing complete suppression of the CDW transition at $x$=0.36 (Supplementary Fig. 1**g**)[23–25].

Apart from the suppression of long-ranged charge orders, the dilute magnetic Cr dopants also introduce local magnetic moments. The local moments are screened by the surrounding itinerant electrons from non-magnetic metallic band of $CsV_3Sb_5$, which results in the Kondo resonance in the density of state[29,30]. The Kondo resonance exhibits a pronounced peak with asymmetric spectral shape near Fermi level in the representative d$I$/d$V$ spectra obtained upon the Cr dopants of Cr-doped $CsV_3Sb_5$ (i.e. $x$ = 0.03, Fig. 2**a**). We define the energy corresponding to the maximum intensity of this resonance system with local electron-hole asymmetry as the Kondo resonance position, $P_{\mathrm{kondo}}$. Such peak is absent from the d$I$/d$V$ spectra of dopant free region. Except for the Kondo resonance peak at $P_{\mathrm{kondo}}$~ -3 meV as shown in Fig. 2**a**, there are no magnetic impurity states inside the superconducting gap, which indicates that the Cr-doped sample reaches the strong coupling regime due to the Kondo screening of impurity magnetic moment[31].

We note that $P_{\mathrm{kondo}}$ could vary from -5 meV to 5 meV for different Cr dopants at Sb surface of Cr-doped sample (Supplementary Fig. 2). The energy range of $P_{\mathrm{kondo}}$ falls well within the range reported in earlier work[36,37]. This variation likely arises from spatial inhomogeneity in the surface electronic states introduced by Cr doping, which modifies the local electronic environment of each magnetic impurity and shifts the Kondo resonance position. The $P_{\mathrm{kondo}}$ variations become more pronounced at higher Cr concentrations (Supplementary Fig. 2) with enhanced inhomogeneity of surface states, further supporting this interpretation.

To advance our understandings on the Kondo resonance, the typical d$I$/d$V$ spectrum has been fitted to the Frota function[32,33]. The half width at half maximum (HWHM) of Kondo peak width fitted by Frota function is determined to be about 1.43 mV (Fig. 2**b**). Based on the fitting of d$I$/d$V$ spectrum obtained at ultra-low temperature (5 mK) and the function $T_K = \Gamma_{HWHM}/3.92k_B$, the Kondo temperature is estimated to be approximately 4.23±0.13 K. The evolution of d$I$/d$V$ spectra with external magnetic field perpendicular to the sample surface ($\boldsymbol{B}_z$) show the splitting of Kondo peak. With increasing $\boldsymbol{B}_z$, the resonance peak first stays mostly unchanged. when $|\boldsymbol{B}_z| > 5$ T, the single resonance peak start split into two peaks at the position of $P_+$ and $P_-$ (Fig. 2**c**). Such evolution is clearer in the second derivative of the d$I$/d$V$ spectra (Fig. 2**d**). The linear splitting of resonance peak corresponds to the Zeeman splitting of spin doublet for the local moment of Cr dopant. We describe the two split peaks with double Frota functions (Supplementary Fig. 3) and obtain the linear relationship between splitting energy $\Delta_{split} = P_+ - P_-$ and magnetic field $B_z$, $\Delta_{split} = 0.93\mu_B|\boldsymbol{B}_z|$ (Fig. 2**e**). The moment estimated from the linear fitting ($0.93\mu_B$) is close to the effective magnetic moment $\mu_{eff} = 1.26 \pm 0.12$ $\mu_B$ per Cr atom[26]. An intuitive explanation is that the spin orientation of local moment around Cr dopant is parallel to the kagome plane, and strong out-of-plane magnetic field flop the spin from in-plane to out-of-plane direction, leading to the Zeeman splitting. Such phenomena is in agreement with Kondo effect in strong coupling regime[34,35].

To estimate the characteristic temperature scale of Kondo resonance, we also study the temperature evolution of d$I$/d$V$ spectra for the same Cr dopant. As the temperature increases from 6 K to 12 K, the resonance peak broadens rapidly with increasing temperature (green dots in Fig. 2**f**). To extract the temperature evolution of Kondo resonance, we apply the Frota model taking the Fermi-Dirac broadening into consideration, that is Hurwitz-Fano lineshape[38] to fit the spectra of each temperature (Fig. 2**g**). We analyze the temperature broadening of the linewidth by employing the empirical expressions $\Gamma_{HWHM} = \sqrt{(\alpha k_B T)^2 + 2\left(k_B T_{K,N}\right)^2}$, yielding the parameter of $T_{K,N} = 11.5 \pm 0.2$ K and $\alpha = 2.10 \pm 0.03$, where the $T_{K,N}$ is the numerical Kondo temperature defined in Ref.[38]. The Kondo temperature is estimated to $T_K = T_{K,N}/2.77 = 4.15 \pm 0.09$ K, comparable to the value of ~4.23 K estimated by the ultra-low temperature single spectrum fitting in Fig. 2**b**.

It is worth noting that the temperature evolution is not sufficiently described by the recently derived expression which is successfully applied in a spin-1/2 Kondo impurity studied in single molecule works [38] as the correlated states in our system may involve other spin interactions insufficiently captured by the model for a $S$=1/2 single magnetic molecule on a metallic substrate. Alternative origins for impurity-induced electronic states near the Fermi level, beyond the Kondo scenario, can be excluded. The spinaron scenario would yield multiple field-sensitive features rather than a single Zeeman-split peak[39], and orbital excitations would produce symmetric step-like inelastic tunneling features[40], none of which are observed.

We next study the real-space distribution of Cr dopant induced Kondo resonance. In contrast to the symmetric real-space distribution of Kondo resonance in the previous reports[41,42], the spatial distribution of Kondo resonance shows symmetry-breaking ripple-like patterns surrounding the Cr dopant. In the STM topographic image obtained on the Sb surface of $x = 0.012$ sample at large sample bias, i.e. 500 mV, the Cr dopants appear as symmetric two-fold dumbbells (Fig. 3**a** and Fig. 1**h**). The d*I*/d*V* spectrum of dopant shown in Fig. 3**a** exhibits a Kondo peak at +5 mV (different from Fig. 2**a** since peak energies vary from dopant to dopant; Supplementary Fig. 2). When the sample bias is set at around the Kondo peak, (i.e. 10 mV in Fig. 3**b** and details see Supplementary Fig. 4), bright "ripples" propagating along one of the lattice directions appear around the dopant (Fig. 3**b**). The distance between each protrusion in the ripple equals to the lattice constant $a_0$.

The bright protrusions are located within the V-formed triangles in kagome structure, the Cr dopant is embedded in the spatial location between the first two bright protrusions. The spatial distribution of Kondo resonance is further demonstrated by the d*I*/d*V* linecut (Fig. 3**c**) along the stripe direction and a direction rotated by 120°, respectively. In addition, the d*I*/d*V* map at 5 meV (Fig. 3**d**), which corresponds to $P_{\text{kondo}}$, exhibit similar symmetry-breaking ripples with low-energy STM image. In contrast, d*I*/d*V* map at energy away from the Kondo resonance, i.e. -5 meV, shows opposite feature with that of 5 meV with stripe pattern formed by a reduction of LDOS (Fig. 3**e**). These observations demonstrate that the anisotropic stripe patterns are characteristic spatial distribution of Kondo resonance. Such anisotropic Kondo resonance is universal with respect to different Cr dopants (same spectroscopic measurement to adjacent Cr dopant is shown in Supplementary Fig. 5). The anisotropic real-space distribution of Kondo resonance is observed on samples with different Cr concentrations, despite the variations of $P_{\text{kondo}}$ (Supplementary Fig. 6). For the local symmetry of a Cr dopant embedded in the bridge site of kagome lattice, there are two mirror symmetry axes. The ripple direction is randomly aligned with one of the four equal lattice directions, breaking all the local mirror symmetries (Supplementary Fig. 7).

External strain is ruled out as the origin of this directionality. Strain from the substrate or mounting would be non-uniform and sensitive to cleaving or mounting conditions, yet the observed directionality is reproducible in the spectra obtained at single Cr dopants across eight samples prepared under varied conditions. Moreover, uniaxial strain in kagome metals is known to produce surface-corrugated regions[43], which are absent in our measurements. Finally, strain-induced anisotropy should lead to distinct strain domains, whereas the ripple directions we observe are randomly aligned along one of the four equivalent lattice axes.

The observed anisotropic Kondo resonance state may originate from either the Kondo screening process or the spatial distribution of Cr dopant-induced magnetic moments. Given the kagome-latticed V-Sb layer, where competing interactions induce magnetic frustration, the anisotropic spatial distribution of magnetic moments is the more plausible explanation. In contrast, the delocalized nature of conduction electrons typically favors isotropic screening. Our density functional theory calculations verify the anisotropic distribution of induced magnetic moments. The kagome-latticed V-Sb layer undergoes a CDW phase transition, where V atoms form alternating triangular and hexagonal arrangements that exhibit slightly different preference for Cr substitution. Specifically, Cr substitution at a hexagonal V site is energetically favored over substitution at a triangular site by 1.6 meV/Cr. The fully relaxed atomic structure of the V-Sb layer with a Cr dopant replacing a hexagonal V atom is shown in Fig. 4**a** and Supplementary Fig. 8**a-b**. The simulated

STM image shown in Supplementary Fig. 8**c** agree with the experimental images. The presence of CDW-induced structural distortion breaks mirror symmetry $M_y$, which passes through the Cr atom and is parallel to the *y-z* plane. We emphasize that the 1Q charge order is not considered in the calculations because it is not correlated with the observed unidirectionality of the Kondo resonance (Supplementary Fig. 7) and is likely of surface origin[44]. Furthermore, Cr substitution introduces in-plane compressive strain, leading to asymmetric Cr-V bond lengths (see Fig. 4**b**), which, in turn, lifts an additional mirror symmetry $M_x$ where the *x-z* plane servers as the mirror plane. Additionally, the comparable lengths of these four Cr-V bonds implies that the Cr dopant can suppress the CDW order, which is consistent with experimental observations where the CDW signal disappears with an increase in Cr concentration (Fig. 1**j**).

The Cr dopant has one more *d* electron than an adjacent V atom, thereby introducing a local magnetic moment of 1.98 $\mu_B$/Cr, surrounded by non-local electrons from neighboring surface V atoms. Due to the magnetic proximity effect, the Cr local moment magnetizes adjacent V atoms through antiferromagnetic (AFM) couplings. Such Cr dopant induced local magnetization leads to strong spin frustration within the kagome V-Sb layer (Fig. 4**c**). This frustration give rise to at least four candidate magnetic configurations (Supplementary Fig. 8**d-g**), each of which contains four AFM and two ferromagnetic (FM) nearest neighbor spin-exchange couplings within the corner-sharing bi-triangle unit show in Fig. 4**c**.

Our DFT calculations reveal that the magnetic configuration shown in Fig. 4**d** (V-Sb layer) is the most stable among these configurations. Regardless of the initial magnetic configuration chosen from those shown in Supplementary Fig. 8**d-g**, the system invariably relaxes to the magnetic configuration illustrated in Fig. 4**d**, where one of the four nearest-neighboring V sites is polarized in alignment with the majority spin orientation (spin-up, in red) of the Cr dopant. The spin-up V and the Cr dopant form a line-like feature in the spin-up density, which extends along the Cr-V direction with gradually diminishing intensity, resulting in a quasi-one-dimensional ripple-like propagation pattern through the lattice (Supplementary Fig. 8**h**). This spin-density distribution in the Cr-V plane induces an anisotropic spin distribution on the Sb surface, as depicted in Fig. 4**e** and Supplementary Fig. 8**i**. These anisotropically distributed spin-up electrons undergo AFM coupling with itinerant electrons at low temperature, leading to the anisotropic Kondo resonance observed in our experiments. The theoretically predicted spatial distribution of spin-up charge density (pink isosurface contour superposing in Fig. 4**e**) at Sb layer is comparable to the contour of experimentally observed Kondo cloud in the d*I*/d*V* mapping acquired near the Fermi level (dotted blue contour in Fig. 4**e**). According to our calculations shown in Supplementary Fig. 9, the FM coupled Cr-V bond consistently shows a moderate bond length (approximately 2.80 Å) among all four, indicating the Cr-V structure locking with spin-exchange interactions rather than 3Q-CDW induced structural variations. The CDW order is further suppressed by Cr doping, the direction of ferromagnetic coupled Cr-V bonds, thus the Kondo resonance direction, is not necessarily related to the initial 3Q-CDW structure. Based on such model, shifting the Cr position step by step along the hexagonal 3Q-CDW bond is expected to rotate the Kondo-screening direction by 60° each time, when considering the global symmetry of hexagonal lattice. This sixfold sequence of orientations is indeed observed in STM images (Supplementary Fig. 10).

The Kondo resonances are observed on all slightly doped sample (Supplementary Fig. 2). With increasing concentrations $x$ of Cr, the resonance peak of representative Kondo resonance gradually becomes broader and lower (Fig. 5**a**). It is worth noting that the peak positions of the Kondo resonance in Fig. 5**a** differ from those in Fig. 2 and Fig. 3, as the d$I$/d$V$ spectra were measured on different Cr dopants (Supplementary Fig. 2). When $x$ increases to 0.09, the Kondo resonance is indistinguishable in the low-energy d$I$/d$V$ spectra (Supplementary Fig. 11). The suppression of Kondo effect with increasing Cr concentrations may arise from enhanced Ruderman–Kittel–Kasuya–Yosida (RKKY) interactions between Cr dopants, which can lead to spin-glass–like behavior and modify the Kondo response through impurity–impurity spin interactions[42,45], as suggested in fully substituted Cr–V kagome systems[46]. However, additional mechanisms may also contribute, and further studies are needed to clarify the underlying origin.

Meanwhile, the superconductivity shows intriguing evolution with Cr concentrations as indicated by the spatially averaged d$I$/d$V$ spectra at a low temperature of 5 mK (Fig. 5**b**). In pristine sample, the superconducting gap show V-shaped pairing gap with residual in-gap states[12]. In the slightly doped sample, the peak-to-peak superconducting (SC) gap size $\Delta_{SC}$ does not intensively change. However, the coherence peak height at the gap edge $P_{SC}$=d$I$/d$V$ ($E$=$\Delta_{SC}$) and SC gap depth $H_{SC}$=$P_{SC}$-d$I$/d$V$($E$=0) show different evolution with $x$. Both $P_{SC}$ and $H_{SC}$ become higher with increasing $x$ and reaches maximum at $x$ = 0.012.

The enhanced coherence peak and SC gap depth indicates higher superfluid density[47]. To further characterize the exotic superconductivity with enhanced gap depth, we apply vertical magnetic fields to the samples with $x$ = 0.007 and 0.021 to investigate possible differences of Abrikosov vortices. By applying 20 mT magnetic field, hexagonal vortex lattice form in both samples (Supplementary Fig. 12). However, d$I$/d$V$ linecut measured across vortex and corresponding second-derivative d$^3I$/d$V^3$ spectra have shown different characteristic for these two different samples. For $x$ = 0.007, an X-shaped spatial evolution of vortex bound state (VBS) is observed (Fig. 5**c**), consistent with well-studied pristine $CsV_3Sb_5$. For $x$ = 0.021, a non-splitting Y-shaped spatial evolution of VBS is observed (Fig. 5**c**), different with pristine $CsV_3Sb_5$ and $CsV_{3-x}Cr_xSb_5$ with $x$ = 0.007. While trivial vortices typically exhibit an X-shaped evolution of VBS, previous studies on $Bi_2Te_3$/$NbSe_2$ heterostructure and 2M-$WS_2$ have identified a non-splitting Y-shaped evolution as a signature of Majorana zero modes[48,49]. The change of the VBS behavior when increasing Cr concentrations from 0 to 0.021 can be regard as a possible modification of topological surface band in superconducting state[50], further supporting the existence of a distinct phase[51].

The phase diagram of Kondo effect and superconductivity with Cr concentrations show the strong interplay of superconductivity with Kondo resonance (Fig. 5**d**). We use the HWHM of the Kondo peaks, extracted using the same Frota fitting procedure as in Fig. 2**b** and applied to multiple dopants at each concentration (upper panel of Fig. 5**d**; see Supplementary Fig. 13 for details), as the assessment of Kondo resonance. Coinciding with the observations of Kondo effect, the superconducting gap size stays mostly unchanged but the gap depth shows a dome like evolution (lower panel of Fig. 5**d**). The gap size is not suppressed by the local moment due to the Kondo screening by the itinerant electrons when the Kondo temperature $T_K$ is higher than $T_c$. The additional role of Cr dopants is the electron doping which may result in the dome-like evolution of gap depth. The enhancement of gap depth indicates a distinct superconductivity phase. Following the disappearance of Kondo effect, both gap size and depth suddenly drop when $x > 0.05$. With Cr concentrations further increase, the suppression of both superconductivity and Kondo effect is observed. In the doping range of $0.05>x>0.012$, despite the continuous increase of the Kondo resonance peak, the superconducting gap depth starts to decrease. Such behavior may arise from additional effects introduced by Cr dopants beyond the formation of local moments, such as disorder-induced suppression of superconductivity[52] or band renormalization effects[53]. Once the doping exceeds a certain level, these mechanisms may become more influential than local-moment physics, resulting in a reduction of the superconducting gap even as the Kondo signature grows.

In conclusion, we show that dilute magnetic doping in a kagome superconductor generates spatially anisotropic Kondo resonances that locally break crystalline symmetries and couple strongly to the superconducting state. By linking frustrated magnetism to symmetry-breaking Kondo screening, our results point to a new route for manipulating superconductivity through local magnetic perturbations. More broadly, our work establishes an impurity-engineering platform to tune the balance between Kondo screening, magnetic interactions and superconducting pairing. Extending this approach through controlled doping, pressure or heterostructure design may enable access to emergent regimes, including tunable Kondo lattices, quantum criticality and potentially topological superconducting phases.

**Acknowledgements:** We thank Dr. Zhan Wang for helpful discussions. The work is supported by grants from the National Natural Science Foundation of China (grant nos. 62488201(H.-J.G.), 92580202 (H.C.), 92477205 (W.J.), 52461160327 (W.J.)), the National Key Research and Development Projects of China (grant nos. 2022YFA1204100 (H.Y. and H.C.), 2023YFA1406500 (W.J.)), the CAS Project for Young Scientists in Basic Research (grant no.YSBR-003 (H.C.)) and the Innovation Program of Quantum Science and Technology (grant no. 2021ZD0302700 (H.-J.G., H.C. and H.Y.)). Z.W. is supported by the US DOE, Basic Energy Sciences (Grant No. DE-FG02-99ER45747) and by Research Corporation for Science Advancement (Cottrell SEED award number 27856). Calculations (W.J.) were performed at the the Physics Lab of High-Performance Computing (PLHPC) and the Public Computing Cloud (PCC) of Renmin University of China.

**Author Contributions:** H.-J. G. and H. C. design the experiments. Z.H., H.C. and H.Z. performed the STM/S experiments and data analysis. Z.Zhao, R.W. and H.Y. prepared the Cr doped $CsV_3Sb_5$ samples. Z.W. did the theoretical consideration. Z.Zhang and W.J. carried out theoretical calculations and analysis. Z.H., H.C., Z.Zhang, W.J. and H.-J.G. wrote the manuscript with input from all other authors. H.-J.G. supervised the project.

## Figure Captions

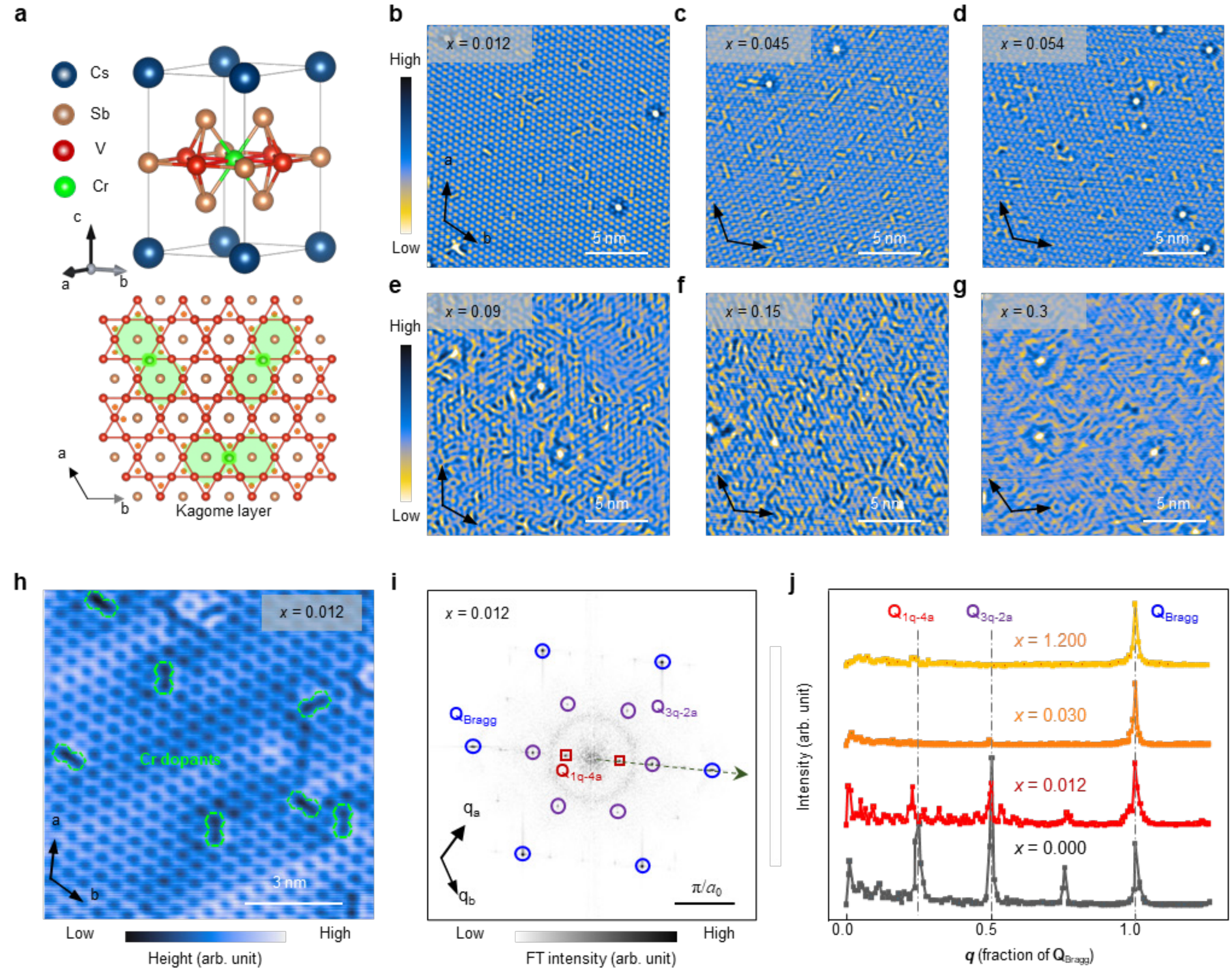

**Fig. 1. Atomic structures of $CsV_{3-x}Cr_xSb_5$ and evolution of CDWs with Cr concentrations**. **a**, Illustration of the $CsV_{3-x}Cr_xSb_5$ crystalline structure. $CsV_{3-x}Cr_xSb_5$ crystallizes in the parent $CsV_3Sb_5$ structure (upper panel). The Cr atoms substitute V atoms in the VSb kagome plane (lower panel). **b-g**, STM images of Sb-terminated surface with different Cr concentrations $x$, obtained at a base temperature of 5 mK. **h**, Atomically-resolved STM image of Sb surface with Cr concentration $x = 0.012$, showing that Cr substitutions appear as two-fold protrusions marked by green dumbbells. ($V_s$ = 500 mV, $I_t$ = 1 nA). **i**, Fourier transform (FT) of STM topography with $x = 0.012$. The Bragg peaks $\mathbf{Q}_{\text{Bragg}}$ are marked in blue circles. The vectors of 2×2 CDW, $\mathbf{Q}_{3q\text{-}2a}$ and $4a_0$ unidirectional charge order $\mathbf{Q}_{1q\text{-}4a}$ are marked in purple circles and red squares, respectively. $V_s$ = 100 mV, $I_t$ = 0.1 nA. **j**, Evolution of FT profile along high symmetry direction (dark green arrow in (**i**)) with Cr concentrations $x$. The peak near $q$=0.75 at $x$=0 results from the satellite peak of $\boldsymbol{Q}_{1q\text{-}4a}$: $\boldsymbol{Q}'_{1Q\text{-}4a} = \boldsymbol{Q}_{\text{Bragg}} - \boldsymbol{Q}_{1Q\text{-}4a}$[28]. The bias of topography for the FT is $V_s$ = -100 mV.

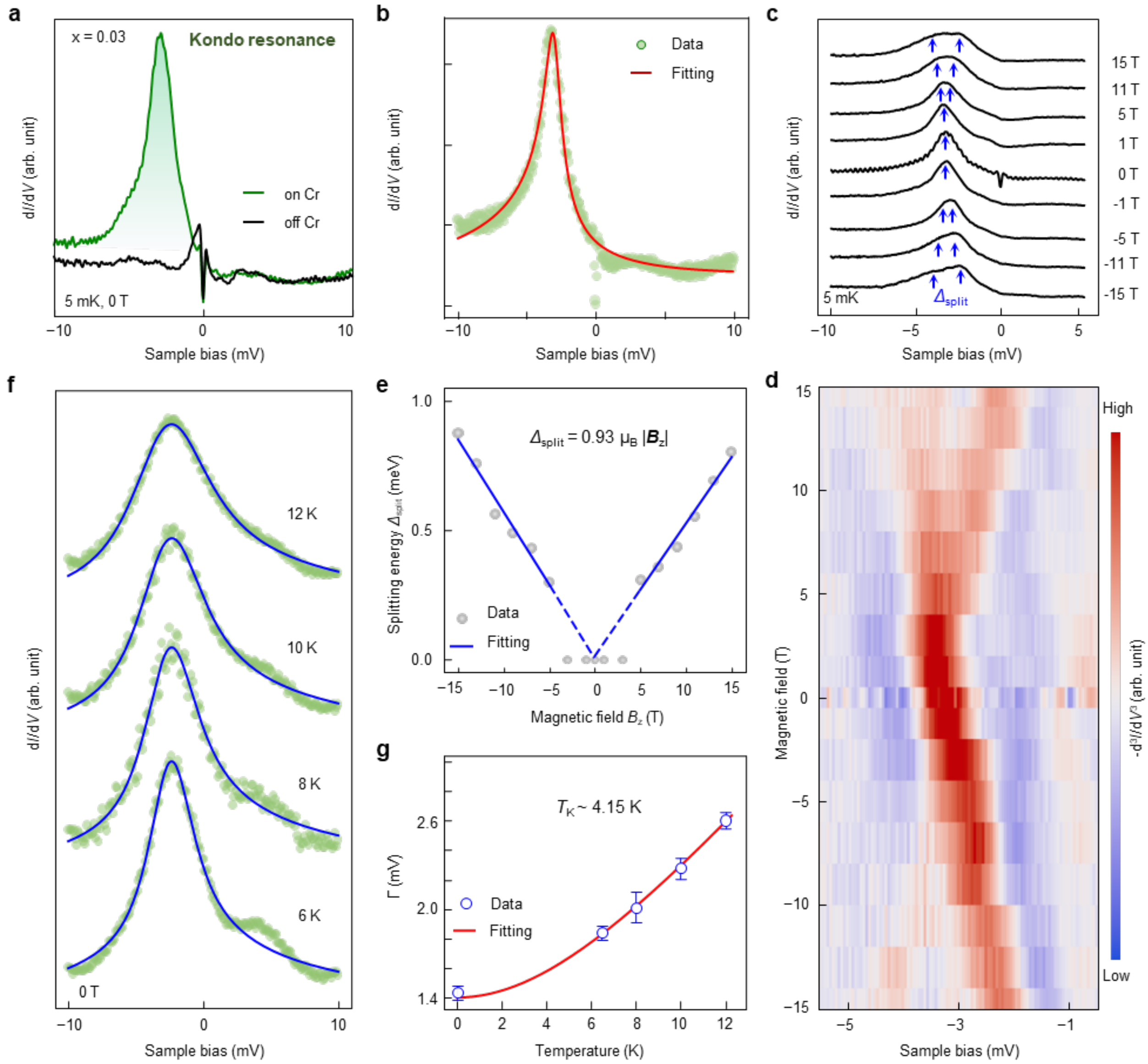

**Fig. 2. Field and temperature evolution of Kondo resonance on the Cr dopants of $CsV_{3-x}Cr_xSb_5$ (x=0.03). a**, d$I$/d$V$ spectra taken away from Cr dopant (black) and on the Cr dopant (green) ($V_s$ = 10 mV, $I_t$ = 1 nA, $V_{mod}$ = 50 μV, $T_{base}$ = 5 mK) showing a pronounced Kondo resonance peak on the Cr dopant. **b**, The d$I$/d$V$ spectra at Cr dopant (green dots), showing that the Kondo peak can be fitted by Frota function (red curve), with a HWHM of ~1.43 mV. **c**,**d** Magnetic field evolution of d$I$/d$V$ spectra (**c**) and the second derivative curves (**d**) on Cr dopants, showing the Zeeman splitting of Kondo peak at field higher than 5 T ($V_s$ = 10 mV, $I_t$ = 1 nA, $V_{mod}$ = 50 μV, $T_{base}$ = 5 mK). **e**, Magnetic field dependence of the splitting Δ (gray dots) with linear fit (blue line), yielding the relation $\Delta_{split} = 0.93\mu_B B$, $\Delta$ shows no sign of splitting with |B| < 5 T and therefore is set to 0. **f**, Temperature dependence of d$I$/d$V$ spectra showing the gradual broadening of the Kondo peak with increasing temperature. Blue curves represent the fitting results by the Frota model taking the Fermi-Dirac broadening into consideration (Hurwitz-Fano lineshape[38]). ($V_s$ = -10 mV, $I_t$ = 1 nA, $V_{mod}$ = 50 μV). **g**, The HWHM of Kondo peak as a function of temperature. The HWHMs are retrieved from the fittings in panel (**f**). The relationship between them is fitted using the empirical expression, which gives α = 2.1 and $T_K$ ~ 4.15 ±0.09 K. Error bars in **g** represent the deviations from the Frota fitting.

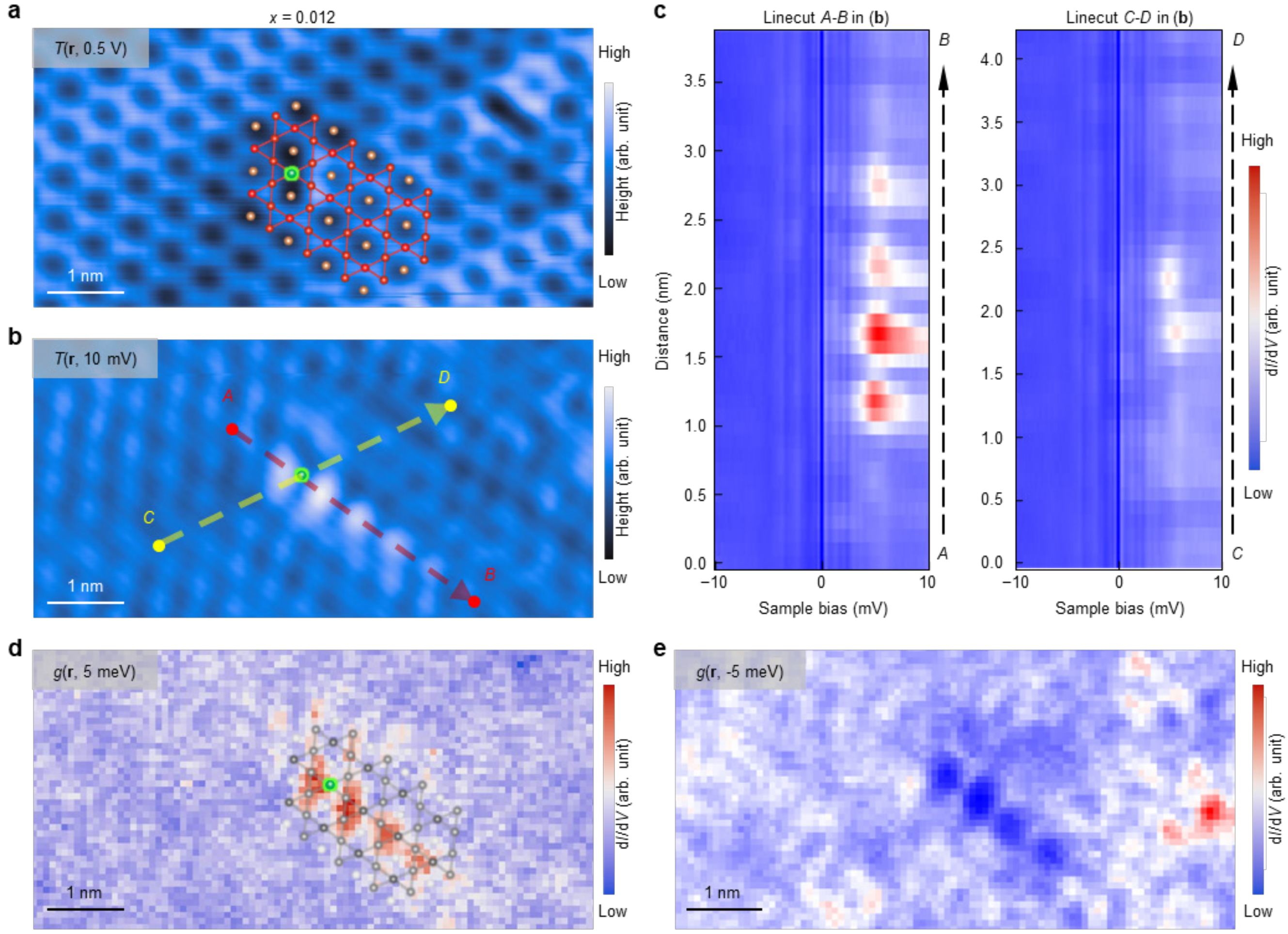

**Fig. 3. The anisotropic distribution of Kondo resonance in the real space. a,b**, Comparison of STM image of the same area with scanning bias of 500 mV (**a**) and 10 mV (**b**) for the sample of $x$=0.012. A schematic model is overlaid in (**a**), showing that the direction of the bright protrusions in (**b**) is along a lattice direction. The distance between each protrusion equals to the lattice constant $a_0$. **c**, Color plot of d$I$/d$V$ linecut across Cr dopant in directions parallel to $a$ axis (red arrow from position "$A$" to "$B$" in (**b**)) and 60° to $a$ axis (yellow arrow from position "$C$" to "$D$" in (**b**)) ($V_s$ = 10 mV, $I_t$ = 1 nA, $V_{mod}$ = 50 μV, $T_{base}$ = 5 mK). **d,e**, LDOS map at 5 meV (**d**) and -5 meV (**e**).

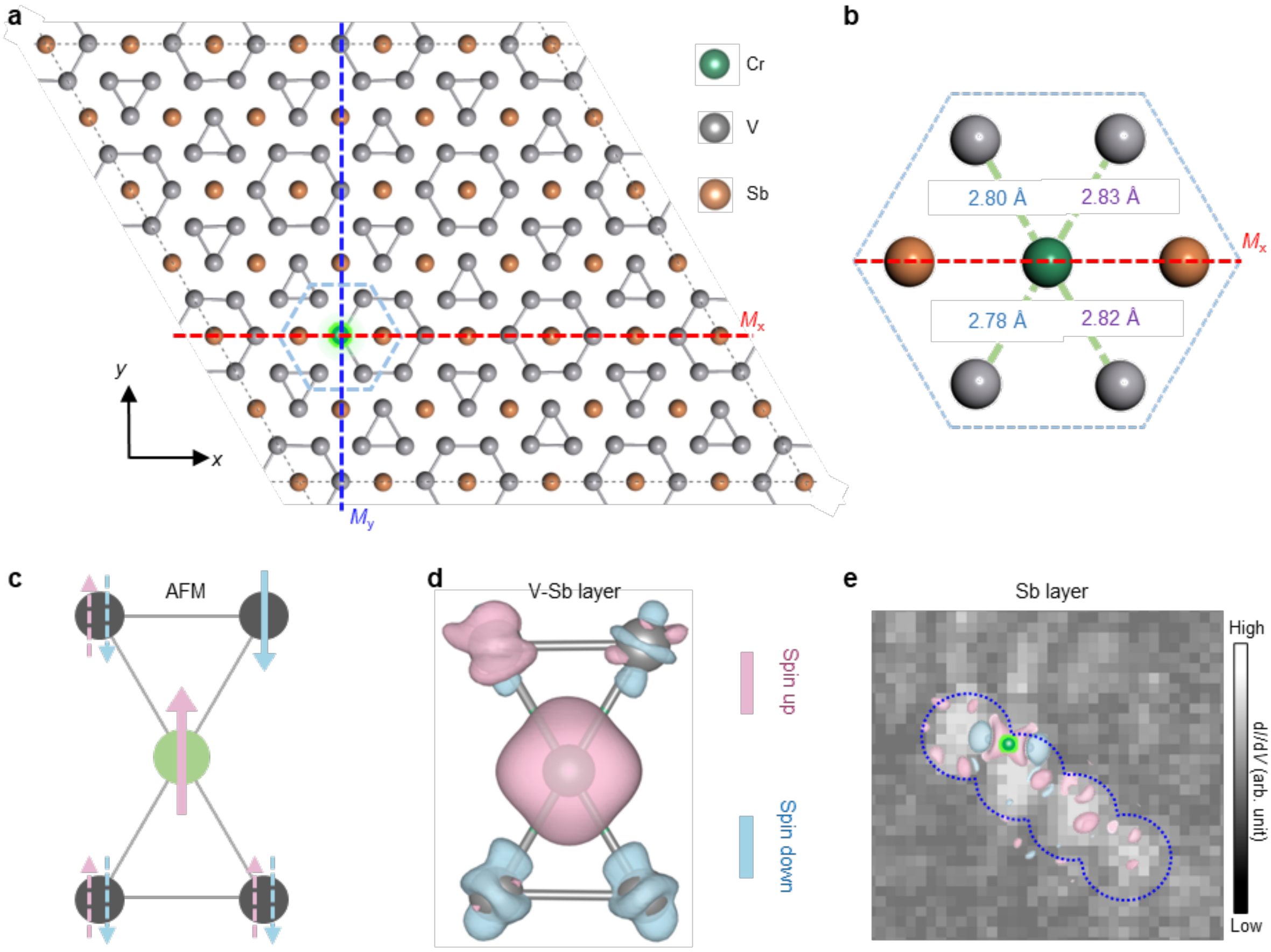

**Fig. 4. DFT calculations of spin density distribution around a Cr dopant in $CsV_3Sb_5$. a,** Fully relaxed atomic structure model of the V-Sb kagome layer after a V atom is substituted by a Cr atom (highlighted by the green shade). **b,** Zoom-in of the blue dashed region in (**a**). Green dashed lines indicate Cr-V bonds, with bond lengths labeled. The blue and orange dashed lines in (**a**) and (**b**) represent the mirror planes for mirror symmetry operations $M_x$ and $M_y$, respectively. If the system has $M_x$ symmetry, two blue (or purple) bond lengths in **b** should be equal. **c,** Schematic of spin frustration in a di-triangle within a kagome lattice where the nearest neighboring sites are coupled antiferromagnetically. The green (black) circle(s) denote the site(s) of the Cr dopant (adjacent V atoms). The violet up and cyan down arrows indicate up- and down-spins, respectively. The pairs of up and down dashed arrows indicate that the spin is undetermined at this site due to spin frustration. **d,** Isosurface contours of spin densities of Cr and its four neighboring V atoms for the most energetically favored magnet configuration as reveal by DFT. The color code follows the same scheme used in panel (**c**) (violet for spin-up and cyan for spin-down). An isosurface value of $2\times10^{-3}$ $e$/Bhor$^3$ was used for plotting. **e,** Experimental d$I$/d$V$ mapping image acquired near a Cr dopant at a bias voltage of 5 mV (black-and-white image), decorated with the theoretical spin density at Sb layer nearby a Cr atom of the magnetic configuration presented in panel **(d),** using the same color scheme in panels **c** and **d**, and an isosurface value of $1.5\times10^{-4}$ $e$/Bhor$^3$. The green bright spot represents the Cr atom, while the dotted blue contour outlines the distribution of the spin-up charge density and the Kondo cloud.

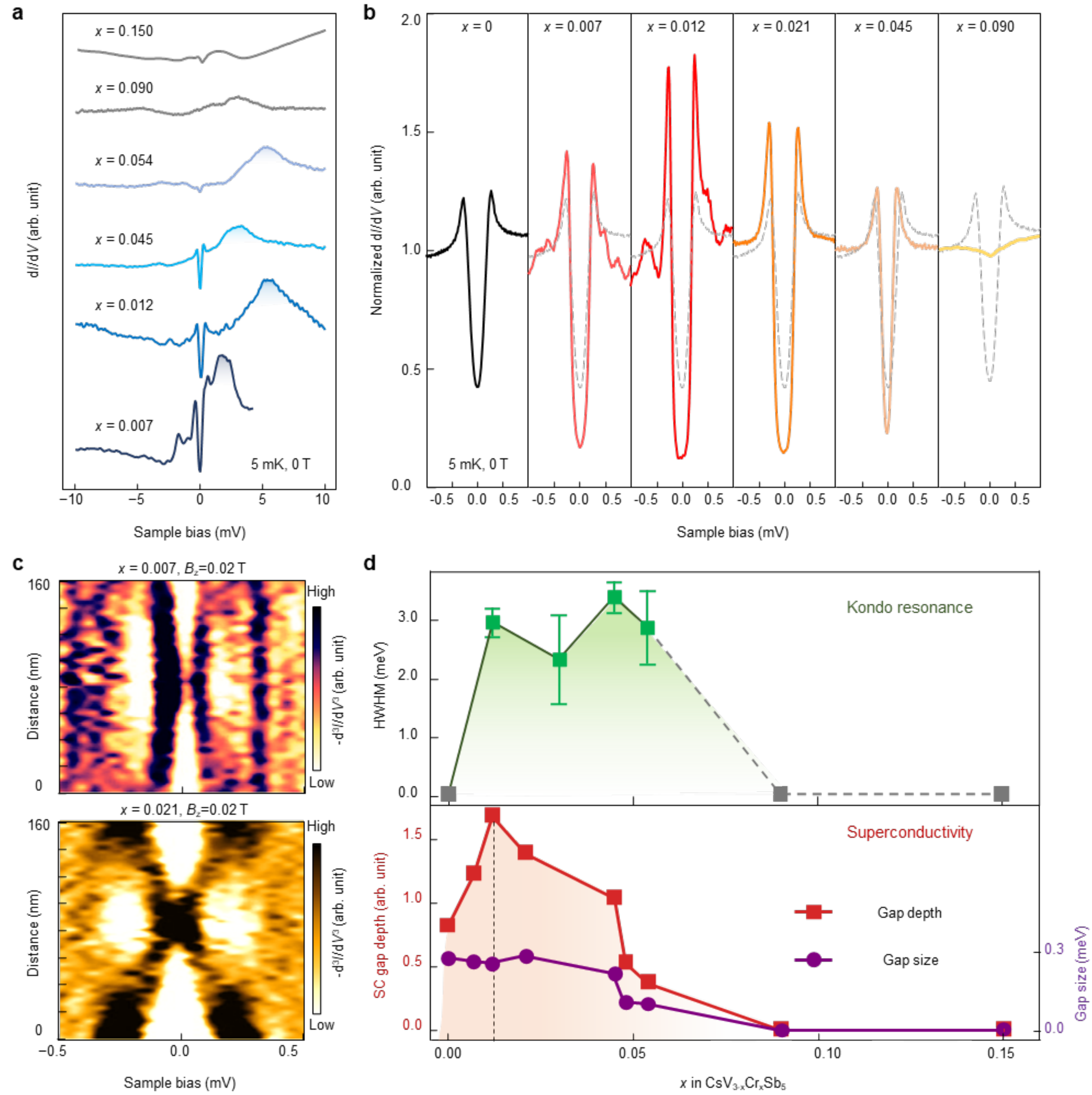

**Fig. 5. Intertwining Kondo resonance with superconductivity in $CsV_{3-x}Cr_xSb_5$. a,** Representative d$I$/d$V$ spectra of Kondo resonance with Cr concentration $x$ = 0.007, 0.012, 0.045,0.054, 0.09 and 0.15, respectively. **b**, Evolution of spatially averaged d$I$/d$V$ spectra with different Cr-concentrations ($V_s$ = -5 mV, $I_t$ = 1 nA, $V_{mod}$ = 10 μV, $T_{base}$ = 5 mK). **c**, Second-derivative plot of the d$I$/d$V$ linecut across the vortex core in sample with $x$ = 0.007 and $x$ = 0.021, showing distinct spatial evolution of vortex bound states. **d,** Evolution of Kondo resonance peak HWHM (up) and superconducting gap size Δ (down) with Cr concentrations $x$, show strong coupling between Kondo screening and superconductivity. The HWHM of Kondo resonance in each concentration are determined by using the same Frota fitting procedure as in Fig. 2**b** and applied to multiple dopants at each concentration. The error bar represents the standard difference of the average HWHM in the fitting of the Kondo resonance for each concentration.

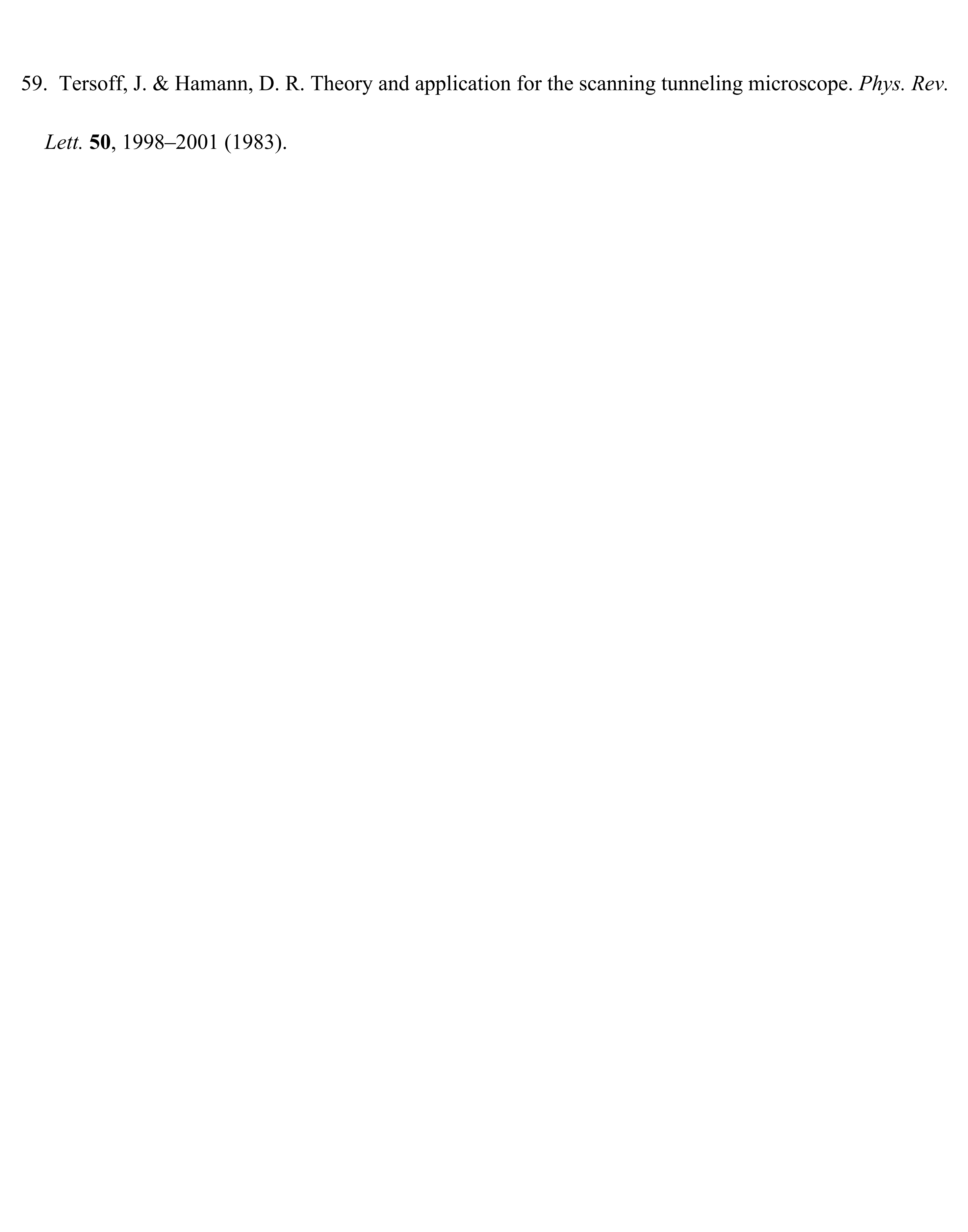

**Methods**

**Single crystal growth of $CsV_{3-x}Cr_xSb_5$**. Single crystals of $CsV_{3-x}Cr_xSb_5$ were grown via a modified self-flux method[12,20].

**Scanning tunneling microscopy/spectroscopy.** The samples used in the STM/S experiments were cleaved at low temperature (10 K) and immediately transferred to an STM chamber and cooled down to 4.2 K. Experiments were performed in an ultrahigh vacuum ($1\times10^{-10}$ mbar) ultra-low temperature STM system equipped with external magnetic field perpendicular to the sample surface. The lowest base temperature is 5 mK with an electronic temperature of 138 mK (calibrated using a standard superconductor, Al crystal. The magnetic field was applied using the zero-field cooling technique. All the scanning parameters (setpoint voltage $V_s$ and tunneling current $I_t$) of the STM topographic images are listed in the figure captions. The d$I$/d$V$ spectra were acquired by a standard lock-in amplifier at a modulation frequency of 973.1 Hz, the modulation bias ($V_{mod}$) is listed in the figure captions. Non-magnetic tungsten tips were fabricated via electrochemical etching and calibrated on a clean Au(111) surface prepared by repeated cycles of sputtering with argon ions and annealing at 500 °C.

**Identification of Cr concentrations.** The diluted Cr concentrations are determined microscopically by counting numbers of dumbbell-like protrusions with high accurate in the STM images of Sb terminated surfaces (Fig. 1**b-d**). With Cr concentrations further increasing to $x > 0.054$, it is difficult to discern the number of dumbbell-like protrusions in the STM images (Fig. 1**e-g**) and therefore the concentrations are verified macroscopically by Energy-Dispersive X-ray Spectroscopy (EDS).

**Density functional theory calculations**. Our density functional theory (DFT) calculations were carried out using the generalized gradient approximation (GGA) for the exchange correlation potential in the form of PerdewBurke–Ernzerhof (PBE)[54], the projector augmented wave method[55], and a plane-wave basis set as implemented in the Vienna ab-initio simulation package (VASP)[56]. We also included the dispersion correction through Grimme's semiempirical D3 scheme[57] in combination with the PBE functional (PBE-D3). A kinetic energy cutoff of 350 eV for the planewave basis was adopted for structural relaxations and electronic structure calculations. A 6×6×2 asymmetric slab model was employed in our calculations. A vacuum layer of 20 Å in thickness was employed to reduce imaging interactions between adjacent supercells. Dipole correction was considered in all calculations to correct the error introduced by the periodic boundary condition and balance the vacuum level differences on the different sides of the polarized surfaces[58]. The lowest layer of the $CsV_3Sb_5$ slab was fixed, while the remaining layers were allowed to relax until the residual force per atom was below 0.01 eV/Å. In the topmost layer, one vanadium (V) atom was replaced by a chromium (Cr) atom to model the substitution of V by Cr. A Gamma-only $k$-mesh was used to sample the Brillouin zone of the super cell. The Methfessel-Paxton method with a $\sigma$ value of 0.05 eV was applied for structural relaxation and electronic structure calculations. The STM images were calculated using the Tersoff-Hamann method[59].